\documentclass{appolb}
\usepackage{graphicx}
\usepackage{slashed}

\begin{document}
\title{Elastic Hadron Scattering in Various Pomeron Models%
\thanks{Presented by P. Erland at XXIII Cracow Epiphany Conference}%
}
\author{\underline{P. Erland$^1$}, R. Staszewski$^2$, M. Trzebi\'nski$^{2,}$\thanks{Corresponding author: maciej.trzebinski@ifj.edu.pl}, R. Kycia$^1$
\address{$^1$ Cracow University of Technology, Warszawska St. 24, 31-155 Cracow\\
$^2$ Institute of Nuclear Physics PAN, Radzikowskiego St. 152, 31-342 Cracow}
}
\maketitle
\begin{abstract}

In this work the process of elastic hadron scattering is discussed. In particular, scattering amplitudes for the various Pomeron models are compared. In addition, differential elastic cross section as a function of the scattered proton transverse momentum for unpolarised and polarised protons is presented. Finally, an implementation of the elastic scattering amplitudes into the GenEx Monte Carlo generator is discussed.

\end{abstract}
  
\section{Introduction}
\begin{figure}[!htbp]
\centering
\includegraphics[width=0.45\textwidth]{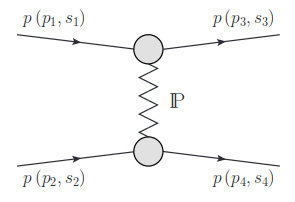}
\caption{Diagram of the elastic scattering: a colour singlet (here a Pomeron) is exchanged between two protons: $p(p_1,s_1)$ and $p(p_2,s_2)$ which are scattered: $p(p_3,s_3)$ and $p(p_4,s_4)$.}
\label{Fey}
\end{figure}

Elastic scattering is the simplest process that one can imagine: in the final state all particles are identical to the initial state ones. This implies that the exchanged object must be a colour singlet and, in particular, that there is no quantum number transfer. In the case of the proton-proton elastic scattering, $pp \rightarrow pp$, see Fig.~\ref{Fey}, such an exchange can be mediated via a photon (electromagnetic interaction) or a Pomeron/Reggeon (strong force).

Elastic scattering is a large fraction of total cross section. However, despite many years of research, there are still open questions concerning its nature.

There is a strong connection between the elastic scattering amplitude and the total cross section, which is described by the optical theorem. The dependence of the total cross section ($\sigma_{tot}$) on the forward scattering amplitude  ($f(\theta=0)$, where $\theta$ is the scattering angle) is given by: $\sigma_{tot}=\frac{4\pi}{k} Im{f(0)}$, where $k$ is the wave vector. This fact is widely used in order to precisely determine the total cross section \cite{ALFA_TOTEM}.

\section{Spin Structure of the Pomeron}
The differential cross section for unpolarized $pp$ elastic scattering is described by~formula:
$$\frac{d\sigma(pp\rightarrow pp)}{dt}=\frac{1}{16\pi s(s-4m_p^2)}\frac{1}{4}\sum_{s_1, \ldots, s_4} | \langle 2s_3,2s_4|\mathcal{T}|2s_1,2s_2 \rangle|^2,$$
where $\langle 2s_3,2s_4|\mathcal{T}|2s_1,2s_2 \rangle$ are the helicity amplitudes with a certain spin orientation of each particle ($s_i$).

Contrary to the photons, the nature of Pomerons is not well known -- there are still many open questions. For example: the Pomeron spin structure.

In the approach of Donnachie and Landshoff, a Pomeron is viewed as a vector object \cite{jeden}. However, as was discussed in \cite{dwa}, such an approach gives a negative x-s value. It is also possible to define it as a scalar or a rank-2 tensor object \cite{dwa}. As was shown in \cite{trzy}, the STAR data \cite{cztery} prefer the tensor over the scalar Pomeron model. 

\subsection{Calculation of the Elastic Scattering Amplitudes}
There are 16 helicity amplitudes describing pp elastic scattering for every combination of spins of incoming and outgoing protons. However, only five of them are independent:
\begin{center}
$\psi_1(s,t)= \langle ++|\mathcal{T}|++ \rangle,$\\
$\psi_2(s,t)= \langle ++|\mathcal{T}|-- \rangle,$\\
$\psi_3(s,t)= \langle +-|\mathcal{T}|+- \rangle,$\\
$\psi_4(s,t)= \langle +-|\mathcal{T}|-+ \rangle,$\\
$\psi_5(s,t)= \langle ++|\mathcal{T}|+- \rangle.$
\end{center}

\noindent$\psi_1$ and $\psi_3$ are the amplitudes describing no spin flip, $\psi_5$ -- single flip, $\psi_2$ and $\psi_4$ -- double flip. These amplitudes can be calculated using a vertex ($\Gamma$) and a propagator ($\Delta$) functions, specific for each Pomeron spin (\textit{cf.} \cite{trzy}):

\begin{itemize}
  \item scalar Pomeron:
  \begin{itemize}
    \item vertex: $i\Gamma^{(I\!\!P_Spp)}(p',p)=-i3\beta_{I\!\!PNN}M_0 F_1[(p'-p)^2]$,
    \item propagator: $i\Delta^{(I\!\!P_S)}(s,t)=\frac{s}{2m_p^2M_0^2}(-is\alpha'_{I\!\!P})^{\alpha_{I\!\!P}(t)}-1$ ,
  \end{itemize}
  \item vector Pomeron:
  \begin{itemize}
    \item vertex: $i\Gamma_{\mu}^{(I\!\!P_Vpp)}(p',p)=-i3\beta_{I\!\!PNN}M_0 F_1[(p'-p)^2]\gamma_{\mu}$,
    \item propagator: $i\Delta_{\mu\nu}^{(I\!\!P_V)}(s,t)=\frac{1}{M_0^2}g_{\mu\nu}(-is\alpha'_{I\!\!P})^{\alpha_{I\!\!P}(t)}-1$,
  \end{itemize}
  \item tensor Pomeron:
  \begin{itemize}
    \item vertex: $i\Gamma_{\mu\nu}^{(I\!\!P_Tpp)}(p',p)=-i3\beta_{I\!\!PNN}F_1[(p'-p)^2]\{\frac{1}{2}[\gamma_{\mu}(p'+p)_\nu+$\\$\gamma_{\nu}(p'+p)_\mu]-\frac{1}{4}g_{\mu\nu}(\slashed p'+\slashed p)\}$,
    \item propagator:\\ $i\Delta_{\mu\nu,\kappa\lambda}^{(I\!\!P_S)}(s,t)=\frac{1}{4}(g_{\mu\kappa}g_{\nu\lambda}+g_{\mu\lambda}g_{\nu\kappa}-\frac{1}{2}g_{\mu\nu}g_{\kappa\lambda})(-is\alpha'_{I\!\!P})^{\alpha_{I\!\!P}(t)-1}$.
  \end{itemize}
\end{itemize}
In these formulas $\beta_{I\!\!PNN}$ is a coupling constant describing the Pomeron-nucleon interaction, $F_1[(p'-p)^2]$ is a form factor, $\gamma_\nu,\gamma_\mu$ are gamma matrices, $\slashed p=\gamma^\mu p_\mu$ is a four momentum in a Feynman slash notation, $\alpha'_{I\!\!P}=0.25$ GeV$^{-2}$ is the Pomeron slope and $\alpha_{I\!\!P}(t)=1.0808+\alpha'_{I\!\!P}$ is the Pomeron trajectory.

The Pomeron spin structure is visible in its propagator formula. For the tensor Pomeron it depends on four variables ($\mu,\nu,\kappa,\lambda$), in contrast to the vector (two variables) and the scalar (no variables) Pomeron models.

\begin{figure}[!htbp]
\centering
\includegraphics[width=0.45\textwidth]{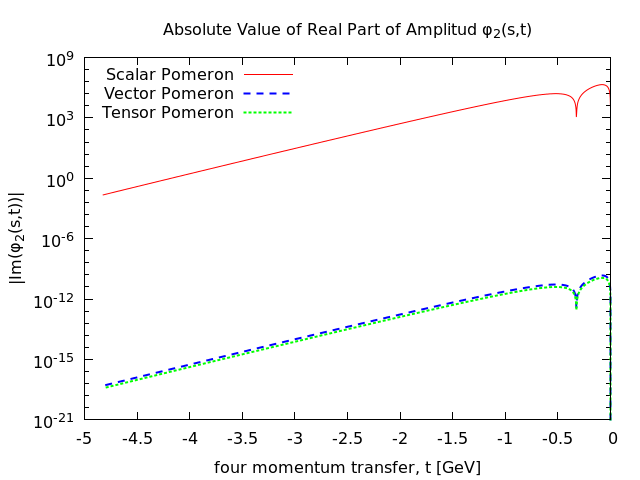}\hfill
\includegraphics[width=0.45\textwidth]{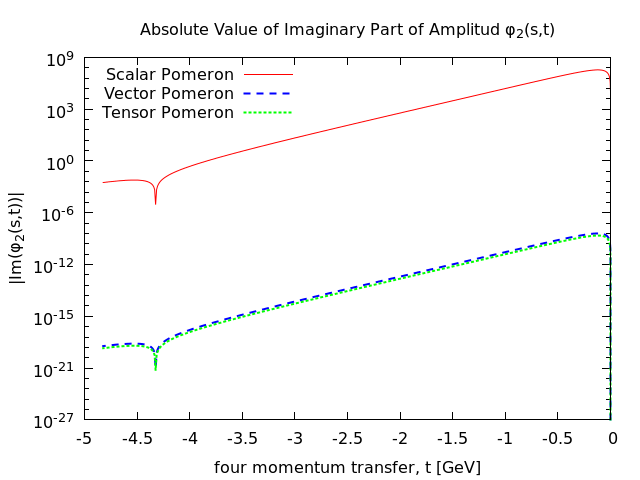}
\caption{Absolute value of real (\textbf{left}) and imaginary \textbf{right} part of $\psi_2$ amplitude. Solid (dashed, dotted) line represent scalar (vector, tensor) Pomeron model.}
\label{ampl}
\end{figure}

The above formulas have been implemented as a set of C++ classes for future implementation in the MC generator. Such approach allows the calculations of more complicated processes to be made in the future. The outcome of an exemplary calculation is shown in Fig. \ref{ampl}, where the absolute value of the imaginary and real part of $\psi_2$ amplitude is plotted for all three discussed Pomeron models.

As can be seen in these figures, the magnitude of the $\psi_2$ amplitude (real and imaginary part) is similar in those of the tensor and vector models, whereas the scalar model predictions are much higher. A dip located close to $t = -0.3$ GeV$^2$ and $t = -4.3$ GeV$^2$ for the real and imaginary part of $\psi_2$ amplitude is due to a change of the sign of the amplitude. The results generated by C++ code were compared with approximate analytic formulas presented in \cite{trzy}. All results are consistent with each other.

\begin{figure}[!htbp]
\centering
\includegraphics[width=0.5\textwidth]{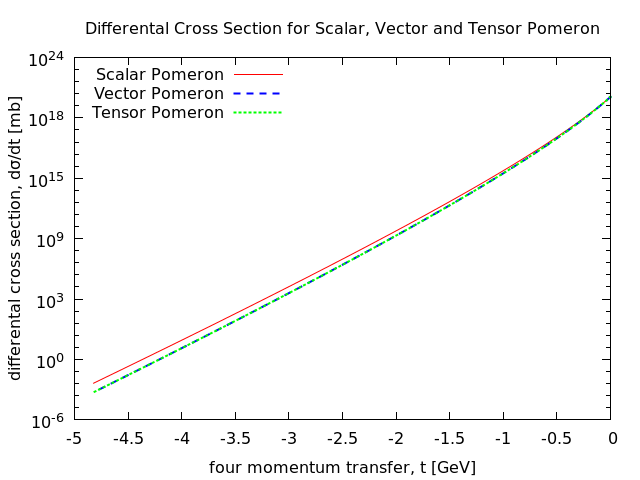}
\caption{Differential cross section given by scalar, vector and tensor Pomeron.}
\label{all_models}
\end{figure}

Since a single amplitude differs a lot between the models, it is interesting to see a cross section integrated over all spin combinations. Results of such calculations are shown in Fig.~\ref{all_models}. For proton-proton collision the tensor (dotted line) and vector (dashed line) Pomeron gives exactly the same results. For small momentum transfers also the scalar model predictions are comparable. They starts to differ (up to a factor of 10) with the increasing value of the four momentum transfer.

\section{Implementation in the GenEx Monte Carlo Generator}
Monte Carlo generators are widely used tools in high energy physics since they provide an essential input helping to understand detector effects. In consequence, they provide a way of comparison between the theory and experimental data. Elastic scattering process is present in many recent HEP MC generators. Based on the formulas described in the previous section, the process of elastic scattering has been added to the GenEx MC generator~\cite{szesc}.

\begin{figure}[htb]
\centerline{\includegraphics[width=0.45\textwidth]{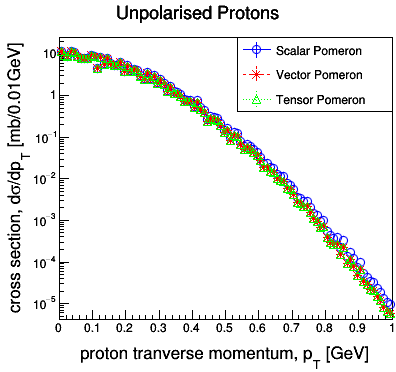}\includegraphics[width=0.45\textwidth]{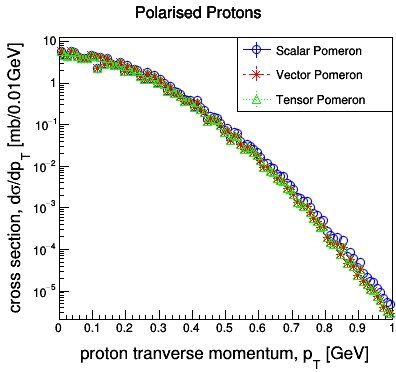}}
\caption{Elastic cross section as a function of scattered proton transverse momentum ($p_T$) for unpolarised (sum of all amplitudes) and polarised beams.}
\label{rozklad1}
\end{figure}

As an example a distribution of the transverse momentum of the final state proton obtained assuming various Pomeron models is shown in Fig.~\ref{rozklad1}. The left plot shows the distribution for the unpolarised protons sum of all amplitudes, whereas the right plot illustrates the polarised (\textit{i.e.} sum of $\psi_1$, $\psi_2$ and $\psi_5$) amplitudes.

For both unpolarised and polarised protons the vector and tensor models are in agreement, whereas the scalar model gives slightly different values for larger transverse momentum values.

\section{Summary and Outlook}
The helicity amplitudes for various Pomeron models for the elastic scattering processes were analysed. It was shown that the differential cross section for the vector and tensor model in proton-proton collisions were in a good agreement, but scalar model differs in region of larger transverse momentum transfer. This difference is also visible in the corresponding, generated MC sample.

An analysis of this simplest possible process gives a good starting point for the future studies of exclusive processes. The plans include further developments of the GenEx generator including the non-resonant and resonant soft exclusive production.

\section{Acknowledgements}
This work was supported in part by Polish National Science Centre grant UMO-2015/17/D/ST2/03530.

\end{document}